\documentclass[epj]{webofc}
\usepackage[varg]{txfonts}   

\newcommand{\be}{\begin{equation}}
\newcommand{\ee}{\end{equation}}
\newcommand{\ba}{\begin{eqnarray}}
\newcommand{\ea}{\end{eqnarray}}

\newcommand{\Tr}{{\rm Tr}}

\woctitle{The Fourth International Workshop on Compound-Nuclear Reactions and Related Topics (CNR*13)}
\begin{document}
\title{Collectivity in Heavy Nuclei in the Shell Model Monte Carlo Approach}
%
%

\author{C. \"{O}zen\inst{1}\fnsep\thanks{\email{cem.ozen@khas.edu.tr}} \and Y. Alhassid\inst{2}\and H. Nakada\inst{3}}

\institute{Faculty of Engineering and Natural Sciences, Kadir Has University, Cibali 34083, Istanbul, Turkey
\and
      Center for Theoretical Physics, Sloane Physics Laboratory, Yale University, New Haven, CT 06520, USA
\and
       Department of Physics, Graduate School of Science,  Chiba University, Inage,  Chiba 263-8522, Japan
}

%

\abstract{%
The microscopic description of collectivity in heavy nuclei in the framework of the configuration-interaction shell model has been a major challenge.  The size of the model space required for the description of heavy nuclei prohibits the use of conventional diagonalization methods. We have overcome this difficulty by using the shell model Monte Carlo (SMMC) method, which can treat model spaces that are many orders of magnitude larger than those that can be treated by conventional methods.
We identify a thermal observable that can distinguish between vibrational and rotational collectivity and use it to describe the crossover from vibrational to rotational collectivity in families of even-even rare-earth isotopes.
We calculate the state densities in these nuclei and find them to be in close agreement with experimental data. We also calculate the collective enhancement factors of the corresponding level densities and find that their decay with excitation energy is correlated with the pairing and shape phase transitions.
}
\maketitle
\section{Introduction}
\label{intro}
  Collective states populate the low-energy spectra of many heavy nuclei and are generally well described by phenomenological models. However, a microscopic description of nuclear
  collectivity within the configuration-interaction shell model approach has been a major challenge.  The use of conventional diagonalization methods has been hampered by the large dimensionality of the many-particle model space. This difficulty can be overcome using an auxiliary-field Monte Carlo method, known in nuclear physics as the shell model Monte Carlo (SMMC) method~\cite{Lang1993,Alhassid1994,Koonin1997,Alhassid2001}.  The SMMC has proven to be a powerful method for the calculation of thermal and statistical properties of nuclei, and in particular level densities~\cite{SMMC-densities,Alhassid2008}

  Here we review recent developments in the applications of SMMC to families of even-even samarium and neodymium isotopes~\cite{Ozen2013a,Alhassid2013a}.
  In particular, we demonstrate, within the framework of a truncated spherical shell model approach, the crossover from vibrational to rotational collectivity in families of isotopes as their number of neutron increases from shell closure towards midshell. We also calculate microscopically collective enhancement factors and study their decay with excitation energy.
\section{The shell model Monte Carlo (SMMC) approach}
\label{sec-smmc}
The SMMC is based on the Hubbard-Stratonovich (HS) transformation~\cite{HS-trans}, in which the Gibbs operator $e^{-\beta H}$ of a nucleus described by a Hamiltonian $H$ at inverse temperature $\beta=1/T$ is represented as a superposition
of one-body propagators of non-interacting nucleons moving in external auxiliary fields $\sigma(\tau)$
\begin{eqnarray}
 e^{-\beta H} = \int D[\sigma] G_\sigma U_\sigma \;.
\end{eqnarray}
Here $G_\sigma$ denotes a Gaussian factor and $U_\sigma$ describes a one-body propagator associated with a given set of auxiliary fields $\sigma$. Using the HS transformation, the thermal expectation value of an observable $O$ at inverse temperature $\beta$ is given by
\begin{eqnarray}
\label{observable}
\langle O\rangle = {\Tr \,( O e^{-\beta H})\over  \Tr\, (e^{-\beta H})} = {\int D[\sigma] W_\sigma \Phi_\sigma \langle O \rangle_\sigma
\over \int D[\sigma] W_\sigma \Phi_\sigma} \;,
\end{eqnarray}
where  $\langle O \rangle_\sigma = \Tr \,(O U_\sigma)/ \Tr\,U_\sigma$ is the thermal expectation value of the observable in a given configuration of the auxiliary fields $\sigma$. Since the number of neutrons and the number of protons are fixed for a given nucleus, all traces in Eq.~(\ref{observable})
are evaluated in the canonical ensemble. Defining a positive-definite function $W_\sigma = G_\sigma |\Tr\, U_\sigma|$ and the associated Monte Carlo sign
$\Phi_\sigma = \Tr\, U_\sigma/|\Tr\, U_\sigma|$, the auxiliary-field configurations $\sigma_k$ are sampled according to $W_\sigma$, and the expectation value in (\ref{observable}) is estimated from
 $\langle  O\rangle \approx  {\sum_k
  \langle O \rangle_{\sigma_k} \Phi_{\sigma_k} / \sum_k \Phi_{\sigma_k}}$.
\section{Collectivity in Heavy Nuclei}
\label{sec-nuccol}
The SMMC approach was shown to be capable of describing the rotational character of $^{162}$Dy---a strongly deformed rare-earth nucleus---in a truncated spherical shell model space~\cite{Alhassid2008}.
Here we discuss recent SMMC applications that extend the study in Ref.~\cite{Alhassid2008} to even-even samarium and neodymium isotopes. Of particular interest is the microscopic description of the crossover from vibrational to rotational collectivity.  The single-particle model space we use consists of the
$0g_{7/2}$, $1d_{5/2}$, $1d_{3/2}$, $2s_{1/2}$, $0h_{11/2}$ and $1f_{7/2}$ proton orbitals, and of the $0h_{11/2}$, $0h_{9/2}$, $1f_{7/2}$, $1f_{5/2}$,
$2p_{3/2}$, $2p_{1/2}$, $0i_{13/2}$, and $1g_{9/2}$ neutron orbitals~\cite{Alhassid2008,Ozen2013a}. The bare single-particle energies were chosen so they reproduce 
the Woods-Saxon energies in the spherical Hartree-Fock approximation. The effective two-body interaction consists of monopole pairing interaction terms for protons and neutrons, and
multipole-multipole interaction with quadrupole, octupole and hexadecupole terms.

\subsection{Crossover from Vibrational to Rotational Collectivity}
\label{sec-cross}
\begin{figure}
\centering
\includegraphics[width=0.9\columnwidth ,clip]{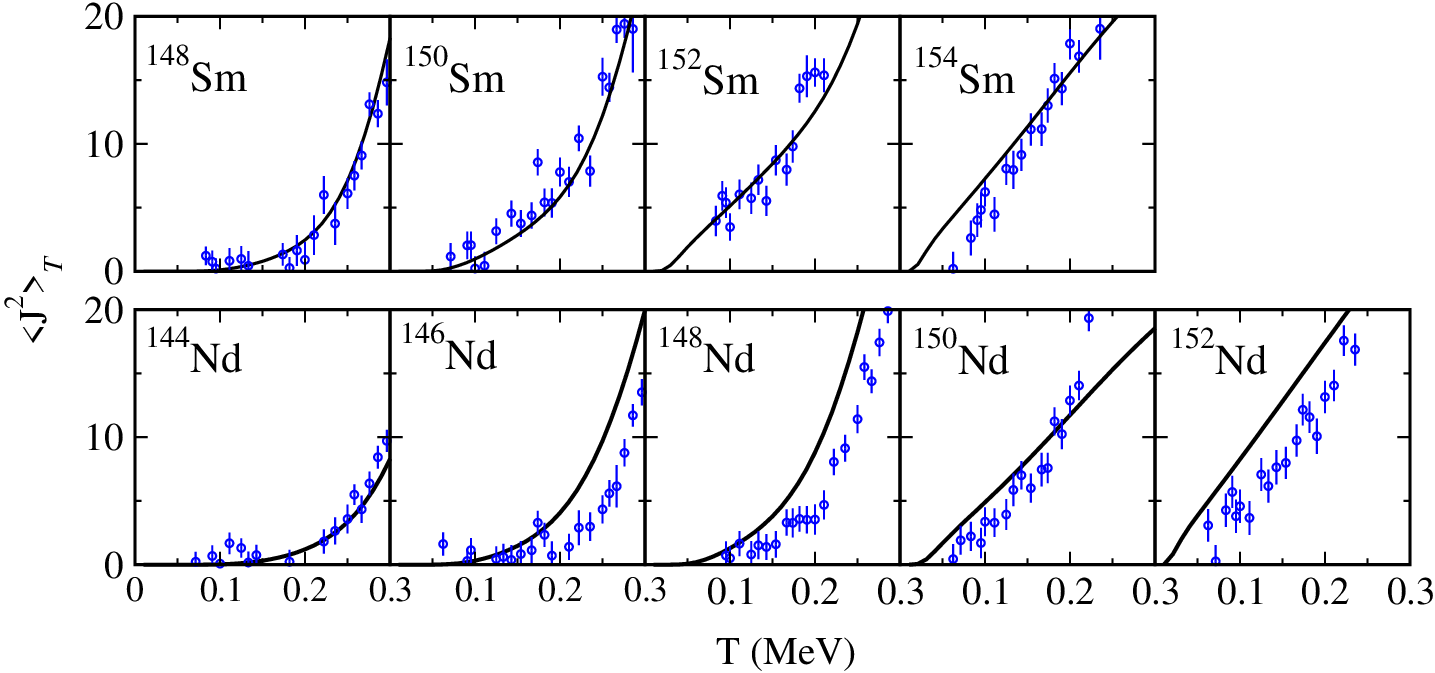}
\caption{$\langle \mathbf{J}^2 \rangle_T$ as a function of temperature in the even-even $^{148-154}$Sm and $^{144-152}$Nd isotopes. The SMMC results (circles with error bars) are
    compared with the experimentally deduced  values obtained using Eq.~(\ref{Eq:J2high}) (for $^{154}$Sm,$^{150}$Nd  and $^{152}$Nd, we use only the discrete sum terms since neutron resonance data are unavailable to determine an experimental BBF state density).
    Adapted from Ref.~\cite{Ozen2013a}. }
\label{fig-J2}
\end{figure}

Collective states are commonly identified through their spectroscopic properties. However, the SMMC approach, as a finite-temperature method, is not suitable for detailed spectroscopic studies. Instead we identify $\langle \mathbf{J}^2 \rangle_T$ as a thermal observable whose low-temperature behavior is sensitive to the type of collectivity (here $\mathbf{J}$ denotes the total angular momentum of the nucleus and $T$ is the temperature at which the expectation value of the observable is evaluated). In Fig.~\ref{fig-J2}, we compare the SMMC results for $\langle \mathbf{J}^2 \rangle_T$  with its experimentally deduced values as a function of temperature for a family of even-mass samarium and neodymium isotopes. These nuclei are known to undergo a phase transition
from spherical shapes (near shell closure) to well-deformed shapes (near mid-shell region) as the number of neutrons increases. Indeed, we observe that for $^{148}$Sm the response of $\langle \mathbf{J}^2 \rangle_T$
to temperature is rather ``soft,'' typical of a vibrational nucleus. With the addition of neutrons, the response evolves gradually to an approximately linear function of temperature in $^{154}$Sm, characteristic of a rotational
nucleus. A similar behavior is observed in the family of neodymium isotopes ranging from $^{144}$Nd to $^{152}$Nd. Thus, the observable  $\langle \mathbf{J}^2 \rangle_T$
can differentiate between vibrational and rotational nuclei and can be used to describe the crossover from vibrational to rotational collectivity. The SMMC results for  $\langle \mathbf{J}^2 \rangle_T$ are in reasonable agreement with the experimentally deduced values of  $\langle \mathbf{J}^2 \rangle_T$  (solid lines). The latter are calculated from
\begin{eqnarray}
\label{Eq:J2high}
 \langle \mathbf{J}^2 \rangle_T & = & \frac{1}{Z(T)} \left(\sum_i^N J_i(J_i+1)(2J_i+1)e^{-E_{i}/T} + \int_{E_{N}}^\infty d E_x \: \rho(E_x) \: \langle \mathbf{J}^2 \rangle_{E_x} \; e^{-E_x/T} \right)\;,
\end{eqnarray}
 where $Z(T)=\sum_{i}^{N} (2J_i+1) e^{-E_i/T} + \int_{E_{N}}^\infty d E_x \rho(E_x) e^{-E_x/T}$ is the experimental partition
function.  The summation terms in the partition function and in Eq.~(\ref{Eq:J2high}) run over a complete set of experimentally known low-lying levels with excitation energies $E_i$ and spins $J_i$  up to an energy threshold $E_N$. The integral
terms account for the contributions of levels with energies above $E_N$, which in the quasi-continuum limit can be described by a state density $\rho(E_x)$. The latter is parametrized by the backshifted Bethe Formula (BBF), whose parameters are determined by a fit to the level counting data at low excitation energies and the neutron resonance data at the neutron separation energy. The quantity $\langle \mathbf{J}^2 \rangle_{E_x}$ is the average value of $\mathbf{J}^2$ at a given excitation energy $E_x$.

At sufficiently low temperatures and for an even-even nucleus, $\langle \mathbf{J}^2 \rangle_T$  can be approximated by~\cite{Alhassid2008,Ozen2013a}
\begin{eqnarray}
\label{Eq:J2-theory}
\langle \mathbf{J}^2 \rangle_T \approx
 \left\{ \begin{array}{cc}
 30  \frac{e^{-E_{2^+}/T}}{\left(1-e^{- E_{2^+}/T}\right)^2} &{\rm vibrational\; band}  \\
 \frac{6}{E_{2^+}} T & {\rm rotational \;band}
 \end{array} \right. \;,
\end{eqnarray}
where  $E_{2^+}$ is the excitation energy of the first $2^+$ level.
Fitting the vibrational and rotational band formulae in Eq.~(\ref{Eq:J2-theory}) to the calculated SMMC values of $\langle \mathbf{J}^2 \rangle_T$ in vibrational and rotational nuclei, respectively,
we extract the $E_{2^+}$ excitation energies in such nuclei and find them to be in good agreement with the experimental values (see Table~\ref{tab:E2}).  These results provide an additional confirmation that our spherical shell-model Hamiltonian can successfully
reproduce the crossover from vibrational to rotational collectivity in the samarium and neodymium nuclei.

\begin{table}[h!]
\centering
\caption{Comparison of the $E_{2^+}$ energies extracted from Eq.~(\ref{Eq:J2-theory}) in SMMC with the experimental values.}
\label{tab:E2}       
\begin{tabular}{llll}
\hline
Nucleus    & collectivity&  $E_{2^+}$ (MeV)  & $E_{2^+}^{exp}$ (MeV)  \\\hline
$^{148}$Sm & vibrational & $0.538 \pm 0.031$ & 0.550 \\
$^{154}$Sm & rotational  & $0.087 \pm 0.006$ & 0.082 \\
$^{144}$Nd & vibrational & $0.702 \pm 0.062$ & 0.697 \\
$^{150}$Nd & rotational  & $0.132 \pm 0.012$ & 0.130 \\
$^{152}$Nd & rotational  & $0.107 \pm 0.006$ & 0.073 \\
\hline
\end{tabular}
\end{table}

\subsection{State Densities}
\label{sec-dens}
The SMMC method has proven to be particularly useful for the calculation of average state densities. The state density
is the inverse Laplace transform of the canonical partition function, and its average can be obtained by evaluating the integral describing the inverse Laplace transform in the saddle-point approximation.

\begin{figure} [h!]
\centering
\includegraphics[width=0.9\columnwidth,clip]{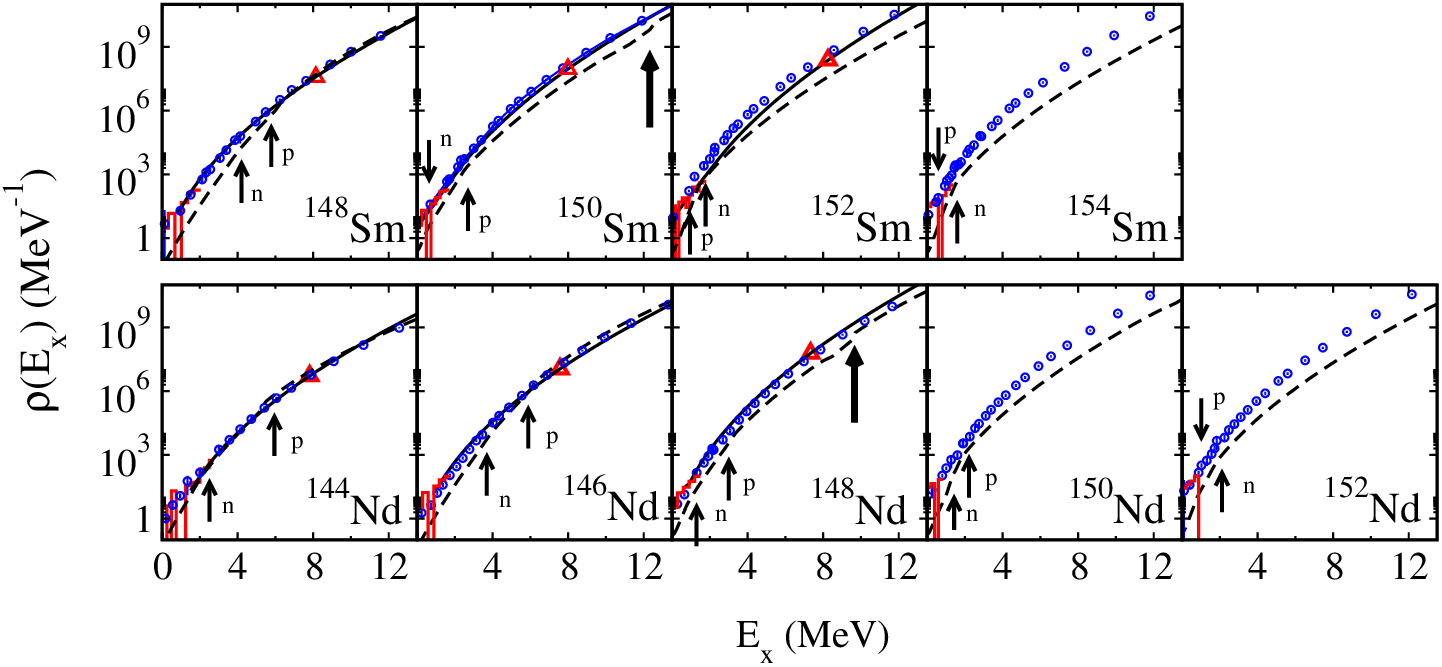}
\caption{State densities in the even-even $^{148-154}$Sm and $^{144-152}$Nd isotopes.
The SMMC densities (circles) are compared with level counting data (histograms) and
neutron resonance data (triangles). The BBF state densities (solid lines), which are determined by a fit to the experimental data, and the HFB densities (dashed lines) are also shown. The arrows indicate the neutron and proton pairing phase transitions, and the thick arrows indicate the shape phase transitions. Adapted from Refs.~\cite{Ozen2013a,Alhassid2013a}.}
\label{fig-rho-even}
\end{figure}

In Fig.~\ref{fig-rho-even} we show the state densities of even-even samarium and neodymium isotopes. The SMMC state densities (circles) are
compared with the experimental state densities obtained from the level counting data at low energies (histograms) and
with the neutron resonance data (triangles) when the latter are available. The solid lines are the BBF state densities, which are determined by a fit to the level counting data at low excitation energies and the neutron resonance data. The dashed lines are the densities obtained from the finite-temperature Hartree-Fock-Bogoliubov (HFB) approximation. As a mean-field approximation, the HFB results provide only the intrinsic states. Thus the difference between the SMMC and the HFB densities arises from
collective bands (vibrational and rotational) that are built on top of the intrinsic states. The ``kinks'' in the HFB density are associated with the neutron and proton pairing phase transitions (arrows) and the shape phase transitions (thick arrows).  $^{148}$Sm, $^{144}$Nd and $^{146}$Nd are spherical in their ground state, hence no shape transition are observed in these nuclei.   Note that the shape phase transitions in $^{152}$Sm, $^{154}$Sm, $^{150}$Nd and $^{152}$Nd occur at higher excitation energies which are not shown in Fig.~\ref{fig-rho-even} (see, however, in Fig~\ref{fig-collective}).

\subsection{Collective Enhancement}
\label{sec-enh}
\begin{figure}
\centering
\includegraphics[width=0.9\columnwidth,clip]{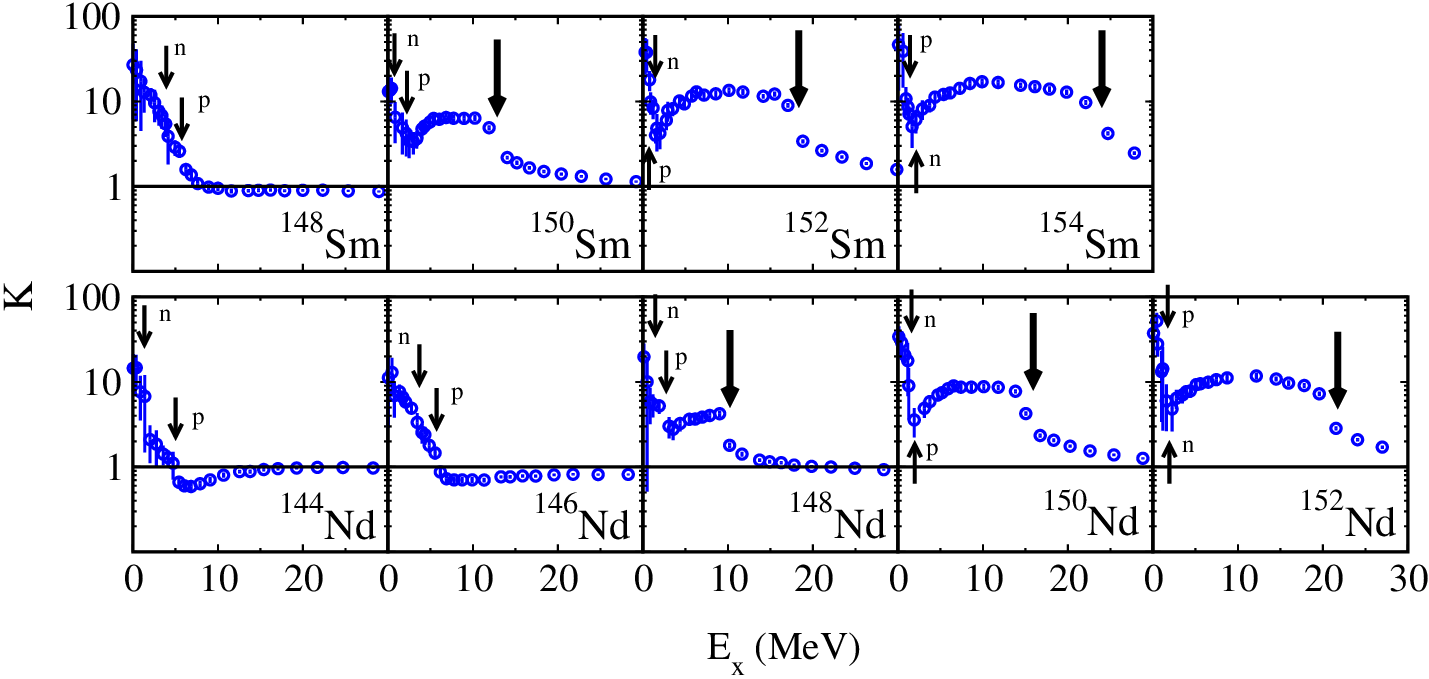}
\caption{Total collective enhancement factor K versus excitation energy $E_x$ in the even-even $^{148-154}$Sm and $^{144-152}$Nd isotopes. The pairing and shape phase transition energies are also shown. Adapted from Refs.~\cite{Ozen2013a,Alhassid2013a}.}
\label{fig-collective}
\end{figure}

The collective enhancement factors account for the collective degrees of freedom in the nuclear state density. The overall collective enhancement factor is usually assumed to factorize into a product of vibrational and rotational enhancement factors, which are often expressed in terms of phenomenological formulae~\cite{RIPL}. Recently we proposed to define microscopically a collective enhancement factor $K$ as the ratio of the SMMC and the HFB state densities, i.e., $K=\rho_\mathrm{SMMC}/\rho_\mathrm{HFB}$~~\cite{Ozen2013a,Alhassid2013a}. In Fig.~\ref{fig-collective}, we show this $K$ as a function of the excitation energy $E_x$ for
the families of samarium and neodymium isotopes. Any collectivity in the spherical nuclei $^{144}$Nd, $^{146}$Nd and $^{148}$Sm should be exclusively vibrational. In these nuclei we observe that collectivity is lost completely (i.e. $K \sim 1$) above the pairing transition energies. However, in deformed nuclei, collective enhancement is due to both vibrational and rotational excitations.
Indeed, collectivity in the deformed nuclei does not disappear above the pairing transitions, and instead $K$ exhibits a local minimum. The persisting collectivity at higher excitation energies, which must be solely rotational, vanishes only above the shape phase transition.

\section{Conclusion}
\label{sec-concl}

We have presented results of recent SMMC studies of the even-even $^{148-154}$Sm and $^{144-152}$Nd isotopes. We have shown that the crossover from vibrational to rotational
collectivity in these rare-earth nuclei can be described microscopically within a truncated spherical shell model space. We have also calculated the total SMMC and HFB state densities and found the SMMC state densities to be in very good agreement with experimental data. We have extracted a microscopic measure of the collective enhancement factor defined by the ratio of the SMMC and HFB state densities.
The damping of vibrational and rotational collectivity is found to be correlated with the pairing and shape phase transitions, respectively.

This work was supported in part by the U.S. Department of Energy Grant No. DE-FG02-91ER40608, and by the Grant-in-Aid for Scientific Research (C) No. 25400245 by the JSPS, Japan. Computational cycles were provided
by the NERSC high performance computing facility at LBL and by the facilities of the Yale University Faculty of Arts and Sciences High Performance Computing Center.




%
%
%

%
\end{document}